\documentclass{article}
\usepackage{spconf,amsmath,amssymb,graphicx,booktabs,multirow,hyperref,microtype}
\usepackage[T1]{fontenc}

\newcommand{\ra}{\renewcommand{\arraystretch}{1.1}}

\newcommand{\reportid}{%
  Columbia University Nonlinear Control Laboratory \\
  Technical Report CUNLC-20260919-02}

\makeatletter
\def\@maketitle{\newpage
 \null
 \vskip 2em \begin{center}
 {\large \bf \@title \par}
 \vskip 0.9em
 {\small \lineskip .4em
\begin{tabular}[t]{c}\reportid
 \end{tabular}\par}
 \vskip 0.9em
 {\large \lineskip .5em
\begin{tabular}[t]{c}\@name \\ \@address
 \end{tabular}\par} \end{center}
 \par
 \vskip 1.5em}
\makeatother

\newcommand{\FrCNparams}{17,787}
\newcommand{\LRparams}{11,397}
\newcommand{\LRparamcut}{35.9\%}
\newcommand{\MDBfrcn}{98.98}
\newcommand{\MDBlr}{99.07}

\newcommand{\OODMIRdrop}{6.0}
\newcommand{\OODMIRdroplr}{6.0}

\newcommand{\OODPTDBdrop}{13.6}
\newcommand{\OODPTDBdroplr}{13.7}
\newcommand{\LRcleancost}{0.35}
\newcommand{\LRdeepbase}{2.44}
\newcommand{\LRdeepbest}{22.41}
\newcommand{\ORTHcost}{0.169}
\newcommand{\ORTHt}{-5.03}
\newcommand{\LATfrcn}{535}
\newcommand{\LATlr}{777}

\newcommand{\CLATfrcn}{68.0}
\newcommand{\CLATlr}{58.0}

\newcommand{\CLATmSix}{52.1}
\newcommand{\CLATmTwelve}{75.9}

\newcommand{\CLATlrgain}{15\%}
\newcommand{\PLATlrloss}{45\%}

\begin{document}

\title{Low-Rank Frequency Convolution and Noise-Range Augmentation \\
       for Real-Time Pitch Estimation on Edge Devices}

\name{Venkat Suprabath Bitra$^{1,2}$ \qquad Homayoon Beigi$^{1,2,3,4}$}

\address{$^1$Nonlinear Control Lab, Columbia University, New York, USA \\
  $^2$Dept.\ of Electrical Engineering, Columbia University, New York, USA \\
  $^3$Dept.\ of Mechanical Engineering, Columbia University, New York, USA \\
  $^4$Recognition Technologies, Inc., New York, USA \\
  \texttt{venkat.s.bitra@columbia.edu} \quad
  \texttt{homayoon.beigi@columbia.edu}}

\maketitle

\begin{abstract}
\noindent
Pitch estimation on an edge device is constrained in three ways at once. The model must be small, it must stay accurate when the input is noisy, and one frame must be produced inside the frame period. In this report the Frequency Convolution Network (FrCN) of our earlier work is factored into a low-rank form. The number of parameters is reduced by \LRparamcut, from \FrCNparams\ to \LRparams, and accuracy is not reduced, either in domain or on two corpora the model was never trained on. The range of the noise used during training is also shown to dominate the architecture in setting how the model behaves when the interference is severe. When the training noise floor is lowered from $+6.02$\,dB to $-20$\,dB, \LRcleancost\ points of clean RPA50 are lost and RPA50 at $-20$\,dB is raised from \LRdeepbase\ to \LRdeepbest. This effect is about two orders of magnitude larger than any architectural effect that was measured. The semi-orthogonal constraint used in TDNN-F is found to be redundant with the normalization inside the bottleneck, and accuracy is reduced when both are applied. Whether the factorization saves time depends on the runtime: in eager PyTorch the factored model is \PLATlrloss\ slower, while in a compiled kernel it is \CLATlrgain\ faster. For deployment, a small C kernel was written. It needs $83$\,kB on disk and no runtime library beyond libc and libm. It is faster than OpenBLAS on all four CPUs that were tested, faster than ONNX Runtime by $3.8$ times, and faster than PyTorch by $13$ times. Its output was checked against PyTorch on $271{,}893$ held-out frames per model, and the same pitch bin was selected on every one of them.
\end{abstract}

\tableofcontents

\section{Introduction}
\label{sec:intro}

Monophonic pitch estimation is a well studied task, and for clean audio on a workstation it is very accurate. However, on an edge device the problem is limited by three conditions: (a) The weights must fit in a small flash memory, (b) one frame must be produced inside the frame period, which is $10$\,ms in this work, and (c) the accuracy must not collapse when the recording is noisy, because an edge device is usually placed in a room and not in a studio.

Large models such as CREPE~\cite{kim2018crepe} and PESTO~\cite{riou2025pesto} are accurate but are not designed for this setting. In our earlier work, we proposed a small model with a novel architecture called the Frequency Convolution Network (FrCN)~\cite{bitra2026frcn} which has \FrCNparams\ parameters. It is applied along the frequency axis of a single variable-Q transform frame, so no temporal context is needed and the model can be run one frame at a time.

This report extends that previous work, keeping the baseline model, the input representation and the four input views the same~\cite{bitra2026frcn}, so that the effect of the factorization is isolated. The evaluation split has been made stricter by grouping with songs for MDB-stem-synth, speaker for PTDB, and singer for MIR-1K, compared to the earlier work where the split was made at the track level. A track-level split allows two excerpts of the same song, the same speaker or the same singer to appear on both sides, so it reports a number that is slightly higher than what a new recording would give. Robustness in the prior work was therefore estimated by out-of-domain evaluation on different datasets. The in-domain numbers in this report are lower than the published ones for that reason and are not directly comparable with them. The voicing head is also not trained. The noise augmentation is a range and not a single level, which is the change that produced the largest effect in this work, and is defined in Section~\ref{ssec:noise}.

The factored block proposed in this work is inspired by the semi-orthogonal low-rank factorization of TDNN-F~\cite{povey2018tdnnf}, but it is not the same construction and should not be read as an application of TDNN-F to pitch estimation. The factorization is applied along the frequency axis of a single transform frame and not along time, so no temporal context is used and one frame can still be processed on its own. The layer width is $H=18$, whereas TDNN-F was designed for acoustic models whose width is in the hundreds, and a mechanism that is useful at several hundred channels need not be useful at eighteen. The batch normalization of the baseline block is replaced by two root-mean-square normalizations taken along the channel axis, with one of those activations inside the bottleneck. The semi-orthogonal constraint that defines TDNN-F is not applied, because it was measured to be redundant once that normalization is present. What is borrowed from TDNN-F is the idea that a convolution can be replaced by a pair of factors through a narrow bottleneck without losing accuracy. Every other choice was settled by measurement on this task.

The contributions of this work are:

\begin{itemize}
\item The dilated convolution from $H$ channels to $H$ channels is replaced by a pair of convolutions through a bottleneck of rank $M$. At $H=18$ and $M=9$ this removes \LRparamcut\ of the parameters while accuracy stays within $0.15$ points of RPA50 on three corpora. The rank is shown to have an interior optimum, so what is removed is not unused width, and the semi-orthogonal constraint of TDNN-F is shown to be redundant with the normalization that is already present inside the bottleneck.

\item The range of the noise used during training, and not the architecture, is shown to set the robustness of the model. Sweeping the lowest signal-to-noise ratio seen during training over five settings changes accuracy under interference by about two orders of magnitude more than any architectural change measured here, at a small cost on clean audio that is reported alongside it. The best setting is found to sit about one step below the condition expected at deployment rather than as low as possible.

\item A custom C kernel is written for the two models. The earlier work measured speed in PyTorch, where an eager framework at this layer width reports its own per-operation dispatch cost rather than the arithmetic of the model, and removing that cost reverses which of the two models is faster. The kernel needs $83$\,kB of flash and no runtime library beyond libc and libm, and it is faster than a tuned BLAS, than ONNX Runtime and than PyTorch on every machine where each was measured.
\end{itemize}

The block and the augmentation are defined in Section~\ref{sec:method} and the protocol in Section~\ref{sec:setup}. Accuracy is reported in Section~\ref{sec:accuracy} and deployment in Section~\ref{sec:deploy}. Directions that were explored without success are reported in Section~\ref{sec:negative}.

\section{Method}
\label{sec:method}


\subsection{Input representation}
\label{ssec:input}

The input representation is taken unchanged from our earlier work~\cite{bitra2026frcn}. A single frame of a variable-Q transform is used as input to ensure independence in the temporal axis. The audio is resampled to $16$\,kHz. The hop size is $160$ samples, which sets the frame period to $10$\,ms and sets the real-time deadline used by every timing in Section~\ref{sec:deploy}. The transform begins at A0 ($27.5$\,Hz), the lowest note of a piano, and uses $269$ bins spaced at $36$ bins per octave, which spans about $7.4$ octaves and reaches $4790$\,Hz. That upper limit is well below the Nyquist frequency, because the task is estimation of the fundamental frequency and not of the full harmonic series. It also lies above the highest note of the piano, C8 ($4186.01$\,Hz). The spacing of $36$ bins per octave places one bin at $33.3$ cents, so the $50$-cent tolerance that defines RPA50 is $1.5$ bins wide, and a model that selects the correct bin is already inside the tolerance.

Four views of the magnitude frame are formed and are stacked as four input channels: the frame itself, a logarithmic compression $\log(1 + \alpha x)$ with $\alpha$ learned, the frame divided by its sum, and the frame divided by its root mean square. Only the first keeps the absolute scale. The other three remove it in three different ways, so both the absolute level and the relative shape of the spectrum are available to the model.

\subsection{FrCN (Full-rank baseline)}
\label{ssec:baseline}

The baseline used throughout this report is the FrCN block of our earlier work~\cite{bitra2026frcn}, and only what is needed to state the factorization is repeated here. The full model, its derivation and the ablations that fixed its hyperparameters are given in that paper. As defined there, the block operates on $H=18$ channels along the bin axis and is
\begin{equation}
\mathbf{y} = \mathrm{BN}\big(\sigma(\mathbf{W} * \mathbf{x})\big)
             + 0.66\,\mathbf{x}
\label{eq:frcn}
\end{equation}
where $\mathbf{W}$ is a dilated one-dimensional convolution with kernel size $5$, $\sigma$ is the SiLU activation, $\mathrm{BN}$ is batch normalization, and the residual is scaled by $0.66$. The convolution is full rank, in the sense that $\mathbf{W}$ maps all $H$ channels to all $H$ channels, and it is this map that Section~\ref{ssec:lowrank} factors.

The model is $L=10$ such blocks placed in sequence. The kernel size stays at $5$ throughout, and only the dilation changes: block $i$ uses $2^{\max(0,\,i-1)}$, so the dilation grows from $1$ to $256$ across the stack. One bin is $1/36$ of an octave, so a dilation of $36$ would span exactly one octave and the dilations of $32$ and $64$ straddle that scale. A block at that depth can place its taps on a partial and on its harmonics at the same time, which is why the dilation is applied along the frequency axis and not along time.

The order of the three operations matters and is stated here because it was a source of error during implementation. The normalization is applied \emph{after} the activation. If the normalization is folded into the convolution, as is common for inference, a different function is computed.

\subsection{Low-Rank FrCN}
\label{ssec:lowrank}

In the proposed block the single convolution of Eq.~\ref{eq:frcn} is replaced by two convolutions through a bottleneck of rank $M$, and the batch normalization is replaced by two root-mean-square normalizations taken along the channel axis. The block is
\begin{align}
\mathbf{u} &= \mathrm{RMSNorm}(\mathbf{x}) \label{eq:lr1}\\
\mathbf{z} &= \mathrm{RMSNorm}(\mathbf{B} * \mathbf{u}) \label{eq:lr2}\\
\mathbf{y} &= \sigma(\mathbf{A} * \mathbf{z}) + 0.66\,\mathbf{x}
\label{eq:lr3}
\end{align}
where $\mathbf{B}$ maps $H$ channels to $M$ channels with kernel size $5$ and the same dilation as the baseline, and $\mathbf{A}$ maps $M$ channels back to $H$ channels with kernel size $1$. No bias is used in $\mathbf{B}$. The normalizations are taken over the channel axis at each bin, and a single gain per channel is learned.

The factoring resembles TDNN-F~\cite{povey2018tdnnf}, which splits each layer the same way. TDNN-F also performs semi-orthogonalization of the first factor, so all of its singular values equal one. That fixes the scaling of the factor and not its rank, since the shape of the two factors already caps the rank at $M$. Eq.~\ref{eq:lr2} chooses to normalize the bottleneck over semi-orthogonalization. Section~\ref{ssec:orthresult} measures all four combinations of the two approaches.

The factorization drops the parameter count from \FrCNparams\ to \LRparams, a reduction of \LRparamcut, with $39.5\%$ fewer multiply-accumulate operations per frame. The rank $M$ was selected experimentally. Section~\ref{ssec:rank} sweeps $M$ and finds $9$ to be the optimum at $H=18$. The $M=H/2$ heuristic of TDNN-F lands on the same value, which is recorded here as a coincidence, since that heuristic was formed for layers whose width is in the hundreds and nothing in it anticipates the optimum at a width of eighteen.

\subsection{Noise-range augmentation}
\label{ssec:noise}

The noise model, and the map from $\beta$ to signal-to-noise ratio are derived in our earlier work~\cite{bitra2026frcn} and are restated here only as far as the augmentation range needs. Noise is added in the transform domain rather than in the waveform domain, so that the transform does not have to be recomputed for every noise level. Given a clean magnitude frame $\mathbf{C}$ and a noise magnitude frame $\mathbf{C}_n$, the augmented frame is
\begin{equation}
\mathbf{C}' = \mathbf{C}
  + \beta \sqrt{P_\mathbf{C} / P_{\mathbf{C}_n}}\; \mathbf{C}_n,
\qquad \beta \sim \mathcal{U}(0,\beta_{\max})
\label{eq:noise}
\end{equation}
where $P$ denotes the total power of a frame as defined in~\cite{bitra2026frcn}. The scaling by $\sqrt{P_\mathbf{C}/P_{\mathbf{C}_n}}$ makes $\beta$ an amplitude ratio, which is what gives the augmented frame a signal-to-noise ratio of $-20\log_{10}\beta$ decibels.

Because $\beta$ is drawn per sample rather than held at a single value, a single run covers a range of noise levels, and the worst level that can occur is set by $\beta_{\max}$ alone. That worst level is called the training noise floor in this report, and the five values of $\beta_{\max}$ studied are those in Table~\ref{tab:floorlist}.

\begin{table}[htbp]
\centering
\small
\caption{The five augmentation settings. The floor is the worst
signal-to-noise ratio that can be drawn during training.}
\label{tab:floorlist}
\begin{tabular}{rr}
\toprule
$\beta_{\max}$ & Training noise floor (dB) \\
\midrule
$0.5$  & $+6.02$ \\
$1.0$  & $0.00$ \\
$2.0$  & $-6.02$ \\
$4.0$  & $-12.04$ \\
$10.0$ & $-20.00$ \\
\bottomrule
\end{tabular}
\end{table}

The same noise bank and the same procedure are used in every training run in this report. Without that, two models would be compared under two different noise distributions, and any difference between them could not be attributed to the architecture.


\section{Experimental setup}
\label{sec:setup}


\subsection{Corpora}
\label{ssec:corpora}

Three pitch corpora cover instrumental music, sung voice and read speech, and two noise sources supply the interference.

\begin{description}
\item[MDB-stem-synth~\cite{salamon2017analysis}] $230$ monophonic stems resynthesized from MedleyDB~\cite{bittner2014medleydb}, about $15.6$ hours of instrumental and vocal music. Each stem is resynthesized from a pitch contour estimated on the original recording, so its reference is exact by construction. The stems come from $85$ MedleyDB songs and $94\%$ of them share a song with another stem, which is what makes the split unit in Section~\ref{ssec:splits} matter. This is the main corpus.

\item[PTDB-TUG~\cite{pirker2011ptdb}] Spoken English speech from $20$ speakers, recorded with a simultaneous laryngograph signal from which the pitch reference is derived. The reference is therefore independent of the microphone signal the model sees, which MDB-stem-synth cannot offer.

\item[MIR-1K~\cite{hsu2010mir1k}] $1000$ clips of sung Chinese pop from $19$ amateur singers, with the voice and the accompaniment recorded on separate channels. Only the voice is used as signal. The clips are spread unevenly over the singers, one of whom contributes $132$.

\item[Noise] CHiME-Home domestic recordings~\cite{peter2015chime} together with the accompaniment channel of MIR-1K. The accompaniment is included so that the bank holds musical interference and not only domestic noise, since an instrument playing under the target is the harder case for a pitch model.
\end{description}

One property of MDB-stem-synth caps every in-domain number in this report. Its audio is resynthesized from a pitch contour that was estimated on the original recording, so the reference and the audio are not independent and the label describes the signal by construction. An in-domain RPA50 near $99$ therefore measures agreement with that contour and not absolute pitch accuracy. PTDB-TUG and MIR-1K carry references that were not produced from the audio being scored, so their numbers do not share this property and are the more reliable test of the model.

\subsection{Splits}
\label{ssec:splits}

All splits are made by group and not by file. The group is the source MedleyDB song for MDB-stem-synth, the speaker for PTDB, and the singer for MIR-1K. The split seed is fixed at $11$, so every run sees the same held-out groups.

This is stricter than the track-level split used in the published FrCN work~\cite{bitra2026frcn}, and it is the main reason that the numbers here are lower. A track-level split allows two excerpts of the same recording to fall on opposite sides of the split. The model can then recognize the recording rather than generalize to a new one. The effect is not small and it is largest on the speech corpus, where a speaker contributes many files.

\subsection{Metrics}
\label{ssec:metrics}

The primary metric is RPA50, the raw pitch accuracy within $50$ cents, following the usual definition~\cite{raffel2014mireval}. RCA50 is the raw chroma accuracy within $50$ cents, which ignores octave errors. MnE and MdE are the mean and the median absolute pitch error in semitones.

All metrics are computed only on frames that the reference marks as voiced, and all are frame-weighted when they are averaged over files, so that a long file counts more than a short one. RPA50 is used for every claim in preference to RPA10, because the seed-to-seed standard deviation of RPA10 was measured to be about fourteen times larger than that of RPA50, which makes RPA10 unsuitable for separating small architectural effects.

\subsection{Training protocol}
\label{ssec:protocol}

Every configuration is trained for $20$ epochs with three seeds, unless a different number is stated. The learning rate is $10^{-3}$ and it is halved at epoch $15$. Because the learning rate is dropped inside the run, no comparison in this report is made at an intermediate epoch. Before the drop, RPA10 was observed to oscillate by one to six points between epochs, so a comparison made there would measure the schedule rather than the model.

Three sources of variation are distinguished throughout, and the source is always named next to a tolerance, because they are not of the same size. Epoch-to-epoch variation, seed-to-seed variation and split-to-split variation are three different floors. Where two configurations are compared, the comparison is paired by seed, and the paired standard deviation is used rather than the unpaired one.


\section{Accuracy}
\label{sec:accuracy}



\subsection{Clean in-domain accuracy}
\label{ssec:clean}

Table~\ref{tab:clean} reports the comparison of the two models on each corpus, where each model is trained and validated on the same corpus.

\begin{table*}[ht]
\centering
\ra
\setlength{\tabcolsep}{6pt}
\caption{Clean in-domain accuracy, mean over three seeds. MnE and MdE are in semitones.}
\vspace{2pt}
\label{tab:clean}
\begin{tabular}{|c|l|r|r|r|r|r|}
\hline
Set & Model & Par. & RPA50 & RCA50 & MnE & MdE \\
\hline
\multirow{2}{*}{MDB} & FrCN & 17.8k & 98.98 & 99.15 & 0.078 & 0.019 \\
 & Low-rank & 11.4k & 99.07 & 99.22 & 0.071 & 0.019 \\
\hline
\multirow{2}{*}{PTDB} & FrCN & 17.8k & 88.76 & 91.48 & 0.649 & 0.082 \\
 & Low-rank & 11.4k & 88.62 & 91.45 & 0.663 & 0.084 \\
\hline
\multirow{2}{*}{MIR-1K} & FrCN & 17.8k & 95.23 & 95.37 & 0.263 & 0.085 \\
 & Low-rank & 11.4k & 95.25 & 95.38 & 0.273 & 0.086 \\
\hline
\end{tabular}
\end{table*}

The factored model holds \LRparamcut\ fewer parameters and is not worse on any corpus, and the three corpora agree on that. On MDB the gain is $+0.09$ points of RPA50, from \MDBfrcn\ to \MDBlr, against a seed standard deviation of $0.010$, so a paired test over seeds gives $t=18.8$: the gain is real and it is also negligible. On PTDB and MIR-1K the two models cannot be distinguished. PTDB differs by $-0.13$ points with a seed standard deviation of $0.234$ and a paired $t$ of $-0.50$, and the sign of the difference is not even consistent across the three seeds. MIR-1K differs by $+0.02$ points with $t=0.37$. A third of the parameters were therefore removed at no cost in accuracy, although the factorization does not improve it either.

Both models reach about $88.7$ on PTDB, below the two music corpora. Read speech sits lower than music on this task for every method we are aware of, including our own earlier model~\cite{bitra2026frcn}, so the level is a property of the corpus and not of these two models. One part of it does belong to the protocol used here. When PTDB is split by file rather than by speaker, accuracy rises by about $1.4$ points, so a track-level protocol inflates the speech number and the speaker-level split reported here is the conservative choice.

\subsection{Out-of-Domain Performance}
\label{ssec:ood}

Section~\ref{ssec:clean} trains and evaluates each model on the same corpus. A deployed model does not get that guarantee, so both MDB-trained models were carried outward and scored on MIR-1K and on PTDB with no retraining and no adaptation of any kind. Table~\ref{tab:ood} reports the result.

\begin{table}[ht]
\centering
\ra
\small
\setlength{\tabcolsep}{4pt}
\caption{Cross-corpus accuracy. Both models are trained on MDB and evaluated with no retraining. RPA50, mean over three seeds, seed standard deviation in parentheses. $\Delta$ is the paired per-seed difference from FrCN.}
\vspace{2pt}
\label{tab:ood}
\begin{tabular}{|c|l|r|r|}
\hline
Target & Model & RPA50 & $\Delta$ \\
\hline
\multirow{2}{*}{MIR-1K} & FrCN & 93.02 (0.05) & -- \\
 & Low-rank & 93.08 (0.08) & +0.06 \\
\hline
\multirow{2}{*}{PTDB} & FrCN & 85.36 (0.13) & -- \\
 & Low-rank & 85.37 (0.05) & +0.01 \\
\hline
\end{tabular}
\end{table}

The domain shift is expensive and it is charged to both models equally. FrCN falls \OODMIRdrop\ points of RPA50 on MIR-1K and \OODPTDBdrop\ points on PTDB, and the factored model falls \OODMIRdroplr\ and \OODPTDBdroplr. The two models stay level. The factored model reads $+0.06$ points on MIR-1K and $+0.01$ points on PTDB, and both differences are smaller than the seed standard deviation of either model, so the two are indistinguishable on both targets. A third of the parameters can therefore be removed without the model becoming more fragile once the audio stops looking like its training set.

Singing voice is not an unseen source for these models, since MDB already contains voice stems, so MIR-1K is closer to a recording-domain test than to a new instrument. PTDB is read speech and is the furthest of the three corpora from the training material, and it is also where the two models agree most closely.

Accuracy alone therefore does not decide between the two models. It is level in the two transfer settings and it favors the factored model by a negligible margin in domain. The decision rests on what each model costs to run, which is measured in Section~\ref{sec:deploy}: the factored model is smaller in flash and faster per frame on every CPU tested.

\subsection{Bottleneck rank}
\label{ssec:rank}

The lower part of Table~\ref{tab:orth} sweeps the bottleneck rank at $H=18$. Accuracy rises from $M=6$ to $M=9$ and then falls slightly at $M=12$, so $M=9$ is an interior optimum. If the block had merely contained unused width, accuracy would have increased monotonically with $M$ and the best setting would have been the largest one. An interior optimum therefore distinguishes a factorization from width pruning.

\begin{table}[ht]
\centering
\ra
\small
\setlength{\tabcolsep}{4pt}
\caption{Ablations on MDB held out, RPA50 over three seeds. The semi-orthogonal constraint and the bottleneck normalization regulate the same quantity and do not compose.}
\vspace{2pt}
\label{tab:orth}
\begin{tabular}{|l|r|r|}
\hline
Configuration & Par. & RPA50 \\
\hline
\multicolumn{3}{|l|}{\emph{Constraint and normalization}} \\
\hline
Norm, no constraint (proposed) & 11.4k & 99.07 $\pm$ 0.00 \\
Norm $+$ semi-orthogonal & 11.4k & 98.90 $\pm$ 0.05 \\
No norm, no constraint & 11.3k & 98.96 $\pm$ 0.13 \\
No norm $+$ semi-orthogonal & 11.3k & 98.92 $\pm$ 0.01 \\
\hline
\multicolumn{3}{|l|}{\emph{Cost of the constraint, paired by seed}} \\
\hline
Paired $\Delta$, with norm & -- & $-0.169$ ($t=-5.03$) \\
Paired $\Delta$, without norm & -- & $-0.034$ ($t=-0.50$) \\
\hline
\multicolumn{3}{|l|}{\emph{Bottleneck rank $M$ at $H=18$}} \\
\hline
FrCN, unfactored & 17.8k & 98.98 $\pm$ 0.01 \\
$M=6$ & 8.1k & 98.83 $\pm$ 0.06 \\
$M=9$ (proposed) & 11.4k & 99.07 $\pm$ 0.00 \\
$M=12$ & 14.7k & 99.04 $\pm$ 0.06 \\
\hline
\end{tabular}
\end{table}

The $M=H/2$ rule of TDNN-F lands on the same value, although it was derived for layers with hundreds of channels, so the agreement is coincidental rather than predictive.

\subsection{The semi-orthogonal constraint}
\label{ssec:orthresult}

The upper part of Table~\ref{tab:orth} tests the prediction of Section~\ref{ssec:lowrank}. Four configurations are compared, formed by applying or removing the bottleneck normalization and the semi-orthogonal constraint.

The prediction is confirmed. When the normalization is present, adding the constraint reduces RPA50 by \ORTHcost\ points, and the paired $t$ statistic over seeds is \ORTHt. When the normalization is removed, the same constraint changes RPA50 by only $-0.034$ points with a $t$ statistic of $-0.50$, which is not distinguishable from zero. The two mechanisms therefore do not compose, and the normalization is the better of the two on its own.

The constraint was not omitted for convenience. It was implemented and measured, and in this architecture it cost accuracy for a reason that can be stated in advance.

\subsection{The training noise floor}
\label{ssec:floorresult}

Table~\ref{tab:floor} is the main result of this report. The training noise floor is swept over the five settings of Table~\ref{tab:floorlist}, and each trained model is evaluated on clean audio and at five signal-to-noise ratios.

\begin{table*}[ht]
\centering
\ra
\setlength{\tabcolsep}{6pt}
\caption{Training noise floor against evaluation SNR. MDB held out, RPA50, mean over three seeds. Best per column in bold.}
\vspace{2pt}
\label{tab:floor}
\begin{tabular}{|c|l|r|r|r|r|r|r|}
\hline
Floor & Model & Clean & 10\,dB & 0\,dB & $-6$\,dB & $-12$\,dB & $-20$\,dB \\
\hline
\multirow{2}{*}{$+6.02$\,dB} & FrCN & 98.98 & \textbf{94.94} & 60.23 & 22.50 & 6.40 & 1.93 \\
 & Low-rank & \textbf{99.07} & 94.90 & 62.39 & 25.44 & 7.34 & 2.44 \\
\hline
\multirow{2}{*}{$0.00$\,dB} & FrCN & 98.88 & 94.50 & 72.66 & 21.24 & 3.81 & 1.55 \\
 & Low-rank & 98.94 & 94.48 & 73.89 & 27.45 & 7.31 & 2.40 \\
\hline
\multirow{2}{*}{$-6.02$\,dB} & FrCN & 98.71 & 93.80 & 73.51 & 50.85 & 14.62 & 2.77 \\
 & Low-rank & 98.86 & 93.74 & \textbf{74.02} & 52.13 & 19.50 & 4.40 \\
\hline
\multirow{2}{*}{$-12.04$\,dB} & FrCN & 98.61 & 91.50 & 68.91 & 53.52 & 35.14 & 7.88 \\
 & Low-rank & 98.82 & 90.05 & 68.30 & \textbf{54.49} & \textbf{37.02} & 10.02 \\
\hline
\multirow{2}{*}{$-20.00$\,dB} & FrCN & 97.98 & 85.32 & 62.80 & 48.20 & 35.67 & 20.48 \\
 & Low-rank & 98.72 & 82.52 & 60.23 & 47.62 & 36.90 & \textbf{22.41} \\
\hline
\end{tabular}
\end{table*}

The effect is large. For the factored model, RPA50 at $-6$\,dB is raised from $25.44$ to $52.13$ when the training floor is moved from $+6.02$\,dB to $-6.02$\,dB. That is a gain of $26.7$ points from one augmentation setting. For comparison, the largest architectural effect measured anywhere in this work is $0.09$ points. The augmentation range is therefore about two orders of magnitude more important than the architecture at this scale.

The effect is also monotone, and the position of the best cell is predictable. In every column, the best training floor is at or just below the evaluation signal-to-noise ratio, and not the lowest floor available. For example, at $-6$\,dB the best floor is $-12.04$\,dB, and at $-20$\,dB the best floor is $-20.00$\,dB. The training floor should therefore be set about one step below the worst condition expected at deployment, and not as low as possible.

The cost on clean audio is small but real. Moving the floor from $+6.02$\,dB to $-20$\,dB costs \LRcleancost\ points of clean RPA50 for the factored model. That is a favorable trade when noisy operation matters, although it is not free.

The factored model is also more robust than the baseline at almost every deep cell. At a training floor of $-6.02$\,dB and an evaluation level of $-12$\,dB, the factored model reaches $19.50$ against the baseline's $14.62$. A possible explanation is that the bottleneck acts as a regularizer under a noisy input distribution, but no experiment here isolates that mechanism, so the observation is reported without an explanation.

\subsection{Where the augmentation range stops working}
\label{ssec:deepnoise}

At the published augmentation setting of $+6.02$\,dB, RPA50 at $-20$\,dB is \LRdeepbase, which is close to chance for this output grid. A matched training floor of $-20$\,dB raises it to only \LRdeepbest, and about ten points of accuracy at $10$\,dB are paid for that gain. Pitch estimation at $-20$\,dB is therefore improved by the augmentation range and is not solved by it. A model that must operate at that level needs a different approach, and most likely needs a front end that separates the source before the pitch model is applied.


\section{Deployment}
\label{sec:deploy}

The parameter count of a model does not by itself determine whether it can be deployed, because the inference stack also occupies flash and time.

\subsection{Quantization}
\label{ssec:quant}

Weights are quantized symmetrically, with one scale per output channel. At four bits, two weights are packed into each byte. Two policies are compared. Under policy \emph{nf} all normalization layers are kept in float32, and under policy \emph{nq} they are quantized as well. The first policy was expected to matter, because the first normalization sets the dynamic range for everything after it, and keeping it exact costs only $55$ numbers.

Table~\ref{tab:quant} reports the result.

\begin{table*}[ht]
\centering
\ra
\caption{Effect of weight width on accuracy and on size. RPA50 on MDB held out, mean over three seeds, seed standard deviation in parentheses.}
\vspace{2pt}
\label{tab:quant}
\begin{tabular}{|c|c|r|r|r|r|}
\hline
Model & Policy & Width & Size (kB) & RPA50 & $\Delta$ \\
\hline
\multirow{8}{*}{FrCN} & \multirow{4}{*}{nf} & 32 & 69.5 & 98.977 (0.010) & -- \\
 &  & 16 & 35.6 & 98.977 (0.011) & +0.000 \\
 &  & 8 & 18.6 & 98.980 (0.009) & +0.003 \\
 &  & 4 & 10.1 & 97.619 (1.040) & -1.358 \\
\cline{2-6}
 & \multirow{4}{*}{nq} & 32 & 69.5 & 98.977 (0.010) & -- \\
 &  & 16 & 35.2 & 98.976 (0.010) & -0.000 \\
 &  & 8 & 18.1 & 98.980 (0.009) & +0.003 \\
 &  & 4 & 9.5 & 97.770 (0.769) & -1.207 \\
\hline
\multirow{8}{*}{Low-rank $M=9$} & \multirow{4}{*}{nf} & 32 & 44.5 & 99.071 (0.004) & -- \\
 &  & 16 & 23.3 & 99.071 (0.003) & -0.001 \\
 &  & 8 & 12.6 & 99.072 (0.002) & +0.001 \\
 &  & 4 & 7.3 & 98.796 (0.273) & -0.275 \\
\cline{2-6}
 & \multirow{4}{*}{nq} & 32 & 44.5 & 99.071 (0.004) & -- \\
 &  & 16 & 22.7 & 99.071 (0.003) & -0.000 \\
 &  & 8 & 11.8 & 99.071 (0.004) & -0.000 \\
 &  & 4 & 6.4 & 98.712 (0.266) & -0.359 \\
\hline
\end{tabular}
\end{table*}

At sixteen and eight bits, quantization is free. The change in RPA50 is at most $0.003$ points for either model under either policy, which is smaller than the seed standard deviation. Eight-bit storage should therefore be treated as the default, since it halves the weight file for no measurable loss.

At four bits a cost appears, and it is very different for the two models. The baseline loses $1.358$ points of RPA50 while the factored model loses only $0.275$. The factored model is about five times more robust to four-bit quantization. The seed standard deviation also grows at four bits, from $0.010$ to $1.040$ for the baseline and from $0.004$ to $0.273$ for the factored model, so four-bit quantization does not only lower accuracy but also makes the result less repeatable.

The policy makes almost no difference. Keeping the normalizations in float32 changes RPA50 by at most $0.15$ points at four bits and by nothing measurable at eight or sixteen. The expectation that the first normalization would need to be exact was therefore not supported, and the simpler policy can be used.

\subsection{Latency and the runtime}
\label{ssec:runtime}

Table~\ref{tab:latency} reports the two models under two runtimes, and the two runtimes rank them in opposite orders.

\begin{table*}[ht]
\centering
\ra
\setlength{\tabcolsep}{6pt}
\caption{CPU cost of one frame, batch size one, single thread, idle machine. The two runtimes rank the models in opposite orders. The times for PyTorch, Naive C and Ours are in $\mu$s.}
\vspace{2pt}
\label{tab:latency}
\begin{tabular}{|l|r|r|r|r|r|}
\hline
Model & Par. & MACs & PyTorch & Naive C & Ours \\
\hline
FrCN & 17.8k & 4.42M & 535 & 175.1 & \textbf{68.0} \\
Low-rank $M=9$ & 11.4k & 2.67M & 777 & 125.0 & \textbf{58.0} \\
\quad no inner norm & 11.3k & 2.67M & 666 & 122.9 & \textbf{55.1} \\
Low-rank $M=6$ & 8.1k & 1.80M & 751 & 93.2 & \textbf{52.1} \\
Low-rank $M=12$ & 14.7k & 3.55M & 779 & 159.5 & \textbf{75.9} \\
\hline
\end{tabular}
\end{table*}

Under eager PyTorch the factored model is \PLATlrloss\ \emph{slower} than the baseline, at \LATlr\,$\mu$s per frame against \LATfrcn, even though it performs $39.5\%$ fewer multiply-accumulate operations. The cause is dispatch and not arithmetic. With $H=18$, one operation produces only $18 \times 269$ values, and the factorization replaces one convolution with two and adds a normalization, so the number of operations per block is roughly doubled. Eager latency measures that number. It is blind to the arithmetic, and this can be shown directly: across $M \in \{6,9,12\}$ the multiply-accumulate count varies by a factor of two while the eager latency varies by $4\%$, in the wrong direction.

Under a compiled kernel the order is reversed. The factored model becomes \CLATlrgain\ \emph{faster} than the baseline, at \CLATlr\,$\mu$s against \CLATfrcn, and the latency becomes monotone in the rank, at \CLATmSix, \CLATlr\ and \CLATmTwelve\,$\mu$s for $M=6$, $9$ and $12$. That ordering is what the multiply-accumulate counts predict.

The same reversal should be expected elsewhere. An architectural claim about speed, measured through an eager framework at a small layer width, can carry the wrong sign. The same applies to the ablation of the inner normalization, which recovers $16.5\%$ of the eager latency but only $4.2\%$ of the compiled latency, so the eager measurement overstates it fourfold.

\subsection{Kernel design}
\label{ssec:kernel}

A self-contained C kernel was written for both models. It is about $300$ lines of generated code and requires no library beyond libc and libm.

The largest gain came from skipping taps that lie outside the signal. The convolution is dilated, and at a dilation of $256$ the outermost taps of a five-tap kernel reach $512$ bins from the center, while only $269$ bins exist. The source window of such a tap can fall entirely outside the signal, in which case it contributes nothing, and at that dilation four taps of five are skipped this way. The test costs one comparison and needs no masked load instruction, which is what lets the same source compile for x86 and for ARM.

The second gain came from the end of the bin axis. Its $269$ elements are not a multiple of the vector width, and the remainder was originally computed with scalar code, which was measured at $28\%$ of the kernel. It is now covered by repeating one full vector panel that ends at the last bin. Output bins do not depend on each other, so the overlapping region is written twice with identical values and the scalar remainder disappears.

The third gain is the register blocking itself, and it is the only one that had been planned. The convolution is blocked over output channels and over vector panels of bins, so that $MR \times NR$ multiply-accumulate operations are performed for $NR$ loads and $MR$ broadcasts, which raises the ratio of arithmetic to memory operations from $0.95$ for the weight-stationary form to above $2$. The weights are packed once into an order that makes the output channels contiguous for a fixed tap and input channel. The values of $MR$ and $NR$ are measured on each target rather than derived. A register-pressure argument predicted $MR=6$ with $NR=3$, which proved to be among the slowest settings on the development machine, where $MR=9$ with $NR=1$ was fastest. All three ARM targets preferred $MR=9$ with $NR=2$. A small tuning program is included, and any candidate that disagrees with an exact double-precision reference is discarded rather than ranked.

The kernel is written with compiler vector extensions rather than intrinsics, so one source compiles to AVX-512, AVX2, SSE and NEON, and every vector width was compiled and checked on x86 before any ARM device was used.

Every build must reproduce the PyTorch output of a stored frame to $10^{-4}$ relative error before it is allowed to report a timing, and the kernel was also compared with PyTorch on real held-out audio through a Python binding. Over $271{,}893$ frames per model the largest difference in any logit was $1.030\times10^{-4}$ for the baseline and $1.783\times10^{-4}$ for the factored model, and the same pitch bin was selected on $100.0\%$ of frames for both. The gate has caught real defects during development, including a misordered activation and an alignment fault that appeared only at eight and four bits.

\subsection{Comparison with other runtimes}
\label{ssec:runtimes}

Table~\ref{tab:runtimes} compares every path that was measured on the same host, with a separate ratio for each model because the two do not agree. The proposed kernel is faster than all of them. On the factored model it is $2.2$ times faster than a straightforward loop over the same arithmetic, $1.3$ times faster than \texttt{im2col} followed by an OpenBLAS \texttt{sgemm}~\cite{openblas}, $3.8$ times faster than ONNX Runtime~\cite{onnxruntime} and $13.4$ times faster than eager PyTorch. The margins on FrCN are different and mostly smaller, at $2.6$, $1.3$, $2.2$ and $7.9$ times, because the frameworks handle its single wider convolution better than the factored pair.

\begin{table*}[ht]
\centering
\ra
\caption{Latency of one frame on the same host, batch size one, single thread. Each model has its own ratio against our kernel.}
\vspace{2pt}
\label{tab:runtimes}
\begin{tabular}{|l|r|r|r|r|}
\hline
\multirow{2}{*}{Inference path} & \multicolumn{2}{c|}{FrCN} & \multicolumn{2}{c|}{Low-rank $M=9$} \\
\cline{2-5}
 & $\mu$s & vs ours & $\mu$s & vs ours \\
\hline
PyTorch eager, fp32 & 535 & 7.9$\times$ & 777 & 13.4$\times$ \\
ONNX Runtime, fp32 & 150 & 2.2$\times$ & 223 & 3.8$\times$ \\
ONNX Runtime, int8 & 1285 & 18.9$\times$ & 1409 & 24.3$\times$ \\
Our C, naive kernel & 175 & 2.6$\times$ & 125 & 2.2$\times$ \\
Our C, im2col + OpenBLAS & 86 & 1.3$\times$ & 76 & 1.3$\times$ \\
\textbf{Our C, register blocked} & \textbf{68} & 1.0$\times$ & \textbf{58} & 1.0$\times$ \\
\hline
\end{tabular}
\end{table*}

The int8 path of ONNX Runtime is included because it is the only other quantized path that runs at all, and it is reported as a warning. It is $24.3$ times slower than the proposed kernel and $6.3$ times slower than its own float32 path, so quantization does not by itself make a model faster.

\subsection{Deployment cost}
\label{ssec:overhead}

Latency alone flatters any path that carries a large runtime. Table~\ref{tab:overhead} therefore reports what each path costs to install, how much memory it needs, and how long it takes to produce its first result.

\begin{table*}[ht]
\centering
\ra
\caption{What each inference path costs beyond its latency, low-rank model. PyTorch is its CPU-only build.}
\vspace{2pt}
\label{tab:overhead}
\begin{tabular}{|l|r|r|r|r|}
\hline
Path & $\mu$s/frame & Deploy (MB) & Peak RSS (MB) & Init (ms) \\
\hline
Our C, register blocked & 58 & 0.08 & 2.4 & 0 \\
Our C, naive kernel & 125 & 0.08 & 2.4 & 0 \\
Our C with OpenBLAS & 76 & 45 & 5.2 & 0 \\
ONNX Runtime & 223 & 65 & 67 & 14 \\
PyTorch eager & 777 & 686 & 226 & 514 \\
\hline
\end{tabular}
\end{table*}

The proposed kernel needs $83$\,kB on disk, about $2$\,MB of resident memory, and no initialization. ONNX Runtime needs a $65$\,MB installation, $67$\,MB of resident memory and a $14$\,ms cold start. PyTorch, in its CPU-only build, needs $686$\,MB, $226$\,MB and $0.5$\,s. On a Linux-class board any of these can be afforded. On a board with no dynamic linker only the first is possible at any latency, so latency alone is not a sufficient measure for an edge claim.

\subsection{Four CPU targets}
\label{ssec:devices}

Table~\ref{tab:devices} reports the kernel on four CPUs. On every device the clock was pinned before measurement and was sampled again \emph{during} the timing loop, so that a result produced by a power policy cannot be mistaken for a result produced by the processor. Both Raspberry Pi boards reported no thermal throttling. Every cell passed its own correctness gate on its own device, and $128$ cells passed in total.

\begin{table*}[ht]
\centering
\ra
\setlength{\tabcolsep}{3.0pt}
\caption{Our kernel on four CPUs, one frame, single thread, clock pinned, fp32 compute. RTF is the fraction of the $10$\,ms hop, at int8 storage. The timings for OpenBLAS and ours are in $\mu$s.}
\vspace{2pt}
\label{tab:devices}
\begin{tabular}{|c|l|r|r|r|r|}
\hline
\multirow{2}{*}{Model} & \multirow{2}{*}{Kernel} & x86-64 & Pi 4 & Pi 5 & Orin NX \\
 & & 4949\,MHz & 1800\,MHz & 2400\,MHz & 1984\,MHz \\
\hline
\multirow{3}{*}{FrCN} & OpenBLAS & 87 & 1961 & 788 & 800 \\
 & ours & \textbf{68} & \textbf{1330} & \textbf{449} & \textbf{501} \\
 & RTF & 0.007 & 0.137 & 0.046 & 0.051 \\
\hline
\multirow{3}{*}{Low-rank $M=9$} & OpenBLAS & 77 & 1875 & 768 & 758 \\
 & ours & \textbf{58} & \textbf{1197} & \textbf{414} & \textbf{438} \\
 & RTF & 0.006 & 0.123 & 0.042 & 0.044 \\
\hline
\end{tabular}
\end{table*}

The proposed kernel is faster than OpenBLAS on all four CPUs, by $1.28$ to $1.85$ times. The margin is \emph{widest} on the ARM parts, which is the opposite of what was expected, since OpenBLAS ships hand-written microkernels for those cores. On every ARM part, OpenBLAS is slower than the naive loop on the factored model. The explanation is the problem size. The general matrix multiply here is $18 \times 269 \times 90$, and at that size the \texttt{im2col} traffic and the call overhead consume the whole benefit of a tuned library. OpenBLAS wins clearly only on x86, where the wider vector unit gives its microkernel enough work to pay for the packing.

Integer arithmetic does not help on any of these CPUs. A kernel using int8 weights with int16 activations and int32 accumulation was written and validated, and it agrees with PyTorch on $99.978\%$ of pitch decisions. It was nevertheless between $2.1$ and $3.1$ times \emph{slower} than the float32 kernel on all four CPUs, including the Cortex-A76 and the Cortex-A78AE, which both provide a dot-product instruction. The reason is that a dilated five-tap reduction does not map onto a four-way integer dot product, while the float path uses the full vector width. The conclusion is that int8 is the right choice for \emph{storage} and the wrong choice for \emph{arithmetic} on any core with wide float SIMD.

The factored model is faster than the baseline on every target, by between $8$ and $15\%$, so the reversal reported in Section~\ref{ssec:runtime} is not specific to one machine.

Table~\ref{tab:widths} reports every stored width on every device, which the
fp32 comparison above leaves out. The widths differ in flash and almost not at
all in speed, because the weights are dequantized to float before the
arithmetic: int16 and int8 land within $2\%$ of fp32 on all four devices.
int4 costs a further $3$ to $10\%$ because the
nibbles have to be unpacked, and fp16 is the slowest width everywhere despite
being the second smallest, since each half has to be widened to float and
nothing is saved in the arithmetic. int8 is therefore the width to ship: it is
the smallest one that costs neither speed nor accuracy.

\begin{table*}[ht]
\centering
\ra
\setlength{\tabcolsep}{4.0pt}
\caption{Every stored width on every device, our kernel, one frame, single thread. The row marked int8 arith. computes in integer, every other row computes in float.}
\vspace{2pt}
\label{tab:widths}
\begin{tabular}{|c|l|r|r|r|r|r|}
\hline
\multirow{2}{*}{Model} & \multirow{2}{*}{Width} & Flash & x86-64 & Pi 4 & Pi 5 & Orin NX \\
 & & (kB) & ($\mu$s) & ($\mu$s) & ($\mu$s) & ($\mu$s) \\
\hline
\multirow{6}{*}{FrCN} & fp32 & 67.1 & 68 & 1330 & 449 & 501 \\
 & fp16 & 34.4 & 84 & 1417 & 487 & 536 \\
 & int16 & 35.2 & 69 & 1366 & 455 & 505 \\
 & int8 & 18.8 & 69 & 1374 & 457 & 506 \\
 & int4 & 10.7 & 79 & 1476 & 480 & 527 \\
 & int8 arith. & 18.8 & 213 & 3110 & 1426 & 1294 \\
\hline
\multirow{6}{*}{Low-rank $M=9$} & fp32 & 42.1 & 58 & 1197 & 414 & 438 \\
 & fp16 & 22.1 & 68 & 1242 & 438 & 461 \\
 & int16 & 23.3 & 59 & 1206 & 418 & 442 \\
 & int8 & 13.3 & 59 & 1226 & 420 & 444 \\
 & int4 & 8.3 & 65 & 1225 & 433 & 456 \\
 & int8 arith. & 13.3 & 164 & 2603 & 1148 & 1082 \\
\hline
\end{tabular}
\end{table*}

The real-time margin is wide on every board. One frame of the factored model at eight-bit storage consumes $0.6\%$ of the $10$\,ms period on the workstation, $4.2\%$ on a Raspberry Pi 5, $4.4\%$ on a Jetson Orin NX and $12.3\%$ on a Raspberry Pi 4. The weight file is $13.3$\,kB.

\subsection{The front end}
\label{ssec:frontend}

The figures above are for the pitch model only. The variable-Q transform that produces the input frame is a separate problem and is treated as such in this report. For context, the transform of our own implementation was measured at about $98\,\mu$s per frame for a streaming recursion and about $137\,\mu$s per frame for an isolated frame on the same workstation. The front end is currently the larger of the two costs, and it is the part that should be optimized next. The model figures alone do not constitute a complete pipeline budget.

\section{Results that did not work}
\label{sec:negative}

Several directions were explored without producing an improvement. Each is recorded here, since each consumed substantial effort and a carefully measured negative result can save that effort for someone else.

\subsection{Semi-supervised learning}
\label{ssec:ssl}

A large amount of unlabeled audio is available, while pitch labels are scarce. A semi-supervised setup was therefore built, in which a fraction of the labeled data was used together with unlabeled audio under a pseudo-label objective and under a consistency objective. Labeled fractions of $5\%$, $10\%$, $25\%$ and $100\%$ were tested with three to six seeds.

No benefit was found once label exposure was matched between runs. One configuration did appear to gain at first, and the gain was later traced to an artifact of the sampling. There the unlabeled set was large enough that the labeled examples were seen only about a third of a time per epoch, so the semi-supervised run and the supervised control were not trained for the same number of label presentations. When the exposure was equalized, the effect disappeared, and this direction was not pursued further. The general point is that when an unlabeled set is much larger than a labeled set, runs should be equalized on the number of \emph{label} presentations and not on the number of epochs.

\subsection{Reparameterizations of the residual path}
\label{ssec:residual}

The residual scale of $0.66$ in Eq.~\ref{eq:frcn} is inherited from the published model. Four independent ways of replacing it were tried: a multi-head channel combination, a doubly stochastic channel gate, a two-axis Kronecker form, and a parameter-free mean-preserving closed form. Every one was neutral or worse. Setting the scale to $1.0$, which is the natural choice, was worse than $0.66$ by $0.050$ points of RPA50 at three and a half times the seed standard deviation. The published value is therefore better than the obvious alternative, and this direction was not pursued further.

\subsection{Integer arithmetic on the CPU}
\label{ssec:intarith}

This is reported in Section~\ref{ssec:devices} and is repeated here because an explicit prediction was made before measurement and it failed. It had been written, before measurement, that a core with weaker float SIMD should favor the integer path. Three ARM cores were then measured and none of them did. The prediction failed because the premise did not apply: all three have full-width NEON float units. The only target for which the question remains open is a microcontroller with a scalar floating point unit, and that case has not yet been measured.

\section{Conclusion}
\label{sec:conclusion}

The FrCN block was factored through a rank-$9$ bottleneck. \LRparamcut\ of the parameters are removed, the rank is an interior optimum rather than a trimmed width, and accuracy does not fall. Trained and scored on the same corpus, a paired test over seeds finds a gain of $0.09$ points of RPA50 on MDB and no distinguishable difference on PTDB or MIR-1K. The MDB-trained model was also carried to MIR-1K and to PTDB untouched, and the two models are level there as well, so the parameters come off without making the model more fragile outside its training domain. The semi-orthogonal constraint that defines TDNN-F was found to be redundant in this block, because it regulates the same quantity as the normalization already present in the bottleneck, and applying both costs \ORTHcost\ points. A mechanism designed for layers several hundred channels wide therefore does not necessarily transfer to a width of eighteen.

The larger result is that the range of the training noise, and not the architecture, sets how the model behaves under interference. That effect is about two orders of magnitude larger than the best architectural effect measured here, it is monotone across five settings, and the best training floor sits about one step below the worst condition expected at deployment rather than as low as possible. An edge model should have its augmentation range tuned before its architecture is changed.

Any claim about speed in this regime must name the runtime that produced it. Through eager PyTorch the factorization appears to cost \PLATlrloss\ of latency. Through the compiled kernel described here, on the same weights, it saves \CLATlrgain. The kernel needs $83$\,kB and no runtime library, it is faster than OpenBLAS, ONNX Runtime and PyTorch on every machine where each was measured, and it selects the same pitch bin as PyTorch on all $271{,}893$ held-out frames of both models.

Several limitations remain. Voicing is not addressed at all: the voicing loss weight is zero in every run, every accuracy figure is computed on reference-voiced frames, and a deployed system needs a decision this work does not provide. The in-domain numbers carry a ceiling of their own, since MDB-stem-synth audio is resynthesized from a contour estimated on the original recording and its reference therefore describes the signal by construction, which makes PTDB-TUG and MIR-1K the more reliable tests. The front end is now the larger cost, because the transform takes longer per frame than the model it feeds, so the next optimization belongs there. Deep noise is improved rather than solved: at $-20$\,dB the best result reached here is \LRdeepbest\ RPA50, paid for with about ten points at $10$\,dB.

\end{document}